\documentclass[sigconf,natbib=true]{acmart}

\AtBeginDocument{%
  \providecommand\BibTeX{{%
    \normalfont B\kern-0.5em{\scshape i\kern-0.25em b}\kern-0.8em\TeX}}}

\makeatletter
\def\@ACM@checkaffil{
    \if@ACM@instpresent\else
    \ClassWarningNoLine{\@classname}{No institution present for an affiliation}%
    \fi
    \if@ACM@citypresent\else
    \ClassWarningNoLine{\@classname}{No city present for an affiliation}%
    \fi
    \if@ACM@countrypresent\else
        \ClassWarningNoLine{\@classname}{No country present for an affiliation}%
    \fi
}

\usepackage[super]{nth}
\usepackage{amsmath}
\usepackage{amsthm}
\usepackage{mathtools}
\usepackage{bm}
\usepackage{dsfont}
\usepackage{graphicx}

\usepackage{caption}
\usepackage{subcaption}
\usepackage{float}
\usepackage{enumitem}
\usepackage{array}
\usepackage{makecell}
\usepackage{hyperref}
\usepackage{multirow}
\usepackage{tabularx}
\usepackage{makecell} 
\usepackage{url}
\usepackage{xcolor}
\usepackage[flushleft]{threeparttable}
\usepackage[ruled,vlined,linesnumbered]{algorithm2e}
\usepackage{algpseudocode}

\DeclareMathOperator*{\argmax}{arg\,max}

\theoremstyle{definition}

\newcounter{todocounter}
\newtoggle{comments}
 \toggletrue{comments} 
\NewEnviron{doweprint}[1]{%
  \iftoggle{#1}{\BODY}{}%
}

\begin{document}


\title{Augmenting Textual Generation via Topology Aware Retrieval}


\author{Yu Wang}
\email{yu.wang.1@vanderbilt.edu}
\affiliation{%
  \institution{Vanderbilt University}
  \city{Nashville}
  \state{TN}
  \country{USA}
}

\author{Nedim Lipka}
\email{lipka@adobe.com}
\affiliation{%
  \institution{Adobe Research}
  \city{Sanjose}
  \state{CA}
  \country{USA}
}

\author{Ruiyi Zhang}
\email{ruizhang@adobe.com}
\affiliation{%
  \institution{Adobe Research}
  \city{Sanjose}
  \state{CA}
  \country{USA}
}

\author{Alexa Siu}
\email{asiu@adobe.com}
\affiliation{%
  \institution{Adobe Research}
  \city{Sanjose}
  \state{CA}
  \country{USA}
}

\author{Yuying Zhao}
\email{yuying.zhao@vanderbilt.edu}
\affiliation{%
  \institution{Vanderbilt}
  \city{Nashville}
  \state{TN}
  \country{USA}
}

\author{Bo Ni}
\email{boni.bnni@gmail.com	}
\affiliation{%
  \institution{Vanderbilt}
  \city{Nashville}
  \state{TN}
  \country{USA}
}

\author{Xin Wang}
\email{xin.wang.1@vanderbilt.com}
\affiliation{%
  \institution{Vanderbilt}
  \city{Nashville}
  \state{TN}
  \country{USA}
}

\author{Ryan Rossi}
\email{ryrossi@adobe.com}
\affiliation{%
  \institution{Adobe Research}
  \city{Sanjose}
  \state{CA}
  \country{USA}
}

\author{Tyler Derr}
\email{tyler.derr@vanderbilt.com}
\affiliation{%
  \institution{Vanderbilt}
  \city{Nashville}
  \state{TN}
  \country{USA}
}
\renewcommand{\shortauthors}{}

\begin{abstract}
Despite the impressive advancements of Large Language Models (LLMs) in generating text, they are often limited by the knowledge contained in the input and prone to producing inaccurate or hallucinated content. To tackle these issues, Retrieval-augmented Generation (RAG) is employed as an effective strategy to enhance the available knowledge base and anchor the responses in reality by pulling additional texts from external databases. In real-world applications, texts are often linked through entities within a graph, such as citations in academic papers or comments in social networks. This paper exploits these topological relationships to guide the retrieval process in RAG. Specifically, we explore two kinds of topological connections: proximity-based, focusing on closely connected nodes, and role-based, which looks at nodes sharing similar subgraph structures. Our empirical research confirms their relevance to text relationships, leading us to develop a Topology-aware Retrieval-augmented Generation framework. This framework includes a retrieval module that selects texts based on their topological relationships and an aggregation module that integrates these texts into prompts to stimulate LLMs for text generation. We have curated established text-attributed networks and conducted comprehensive experiments to validate the effectiveness of this framework, demonstrating its potential to enhance RAG with topological awareness.
\end{abstract}




\maketitle
\section{Introduction}\label{sec-introduction}
Text-to-text generation, often simply referred to as "text generation", focuses on creating human-readable texts based on the input texts along with task-specific instructions to serve various applications~\cite{garbacea2020neural, gatt2018survey, iqbal2022survey, wang2020towards}, including dialogue systems, document summarization, and academic writing~\cite{rambow2001natural, peng2020few, gupta2010survey, limpo2013modeling}. Despite the unprecedented success achieved by large language models (LLMs) in text generation~\cite{touvron2023llama, brown2020language}, their performance can still be hampered by two key issues: the limited knowledge available in input texts~\cite{yu2022survey, xing2017topic, zhou2018commonsense} and the tendency of LLMs to produce non-factual responses~\cite{zhang2023siren, zhang2023language}. On one hand, relying solely on input text provides limited information misaligning with the abundant knowledge necessary for the desired output. On the other hand, while pre-training over vast corpora has equipped LLMs with world knowledge, this knowledge is encoded in their black-box-like parameters and there is no clear pathway to map these intriguing parameters to interpretable knowledge that can be faithfully used in text generation, which renders the hallucination in the generated responses.

\begin{figure}[t!]
    \centering
    \includegraphics[width=0.5\textwidth]{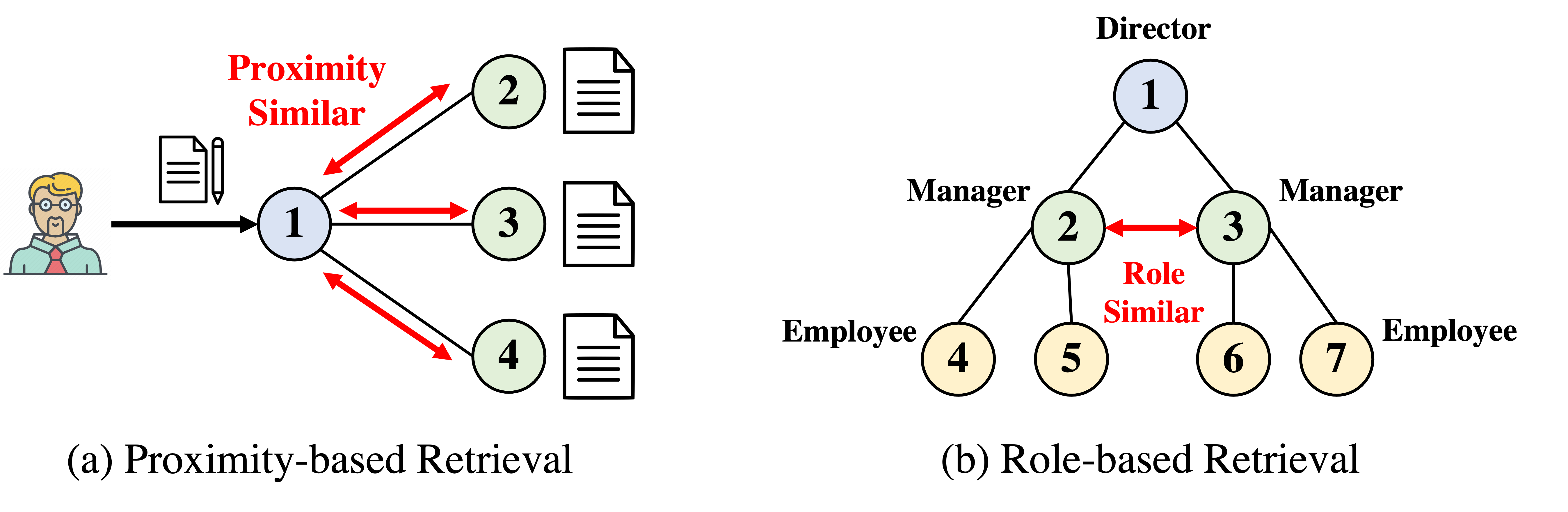}
    \caption{Topology-aware Retrieval for Text Generation.\textbf{(a)}: People write paper abstracts by referring to other papers cited in the related work.; \textbf{(b)}: Employees owning the same local subgraph structures possess the same titles/responsibilities.}
    \label{fig-example}
    \vspace{-4ex}
\end{figure}

To address the challenge of limited knowledge in input texts and avoid hallucination of LLMs, Retrieval-augmented Generation (RAG) can be naturally used to create a well-informed context by retrieving necessary knowledge from external databases~\cite{jin2023large, lewis2020retrieval, mao2020generation, gao2023retrieval}. One notable direction is to improve RAG by integrating additional knowledge into the retrieval process, such as enabling LLMs to utilize automated tools~\cite{zhuang2023toolchain, hao2023toolkengpt, schick2023toolformer} and access well-curated databases~\cite{wang2023knowledge}. For example, \cite{tian2023graph, wang2023knowledge, pan2023unifying} incorporate the symbolic knowledge in the form of structured triples from the knowledge graph to 1) enhance the faithfulness of LLMs' answers, 2) enable the knowledge evolution via dynamic editing, and 3) provide interpretability for LLMs' responses. Unfortunately, most of these KG-enhanced RAGs have been exclusively deployed for question-answering tasks~\cite{tian2023graph, wang2023knowledge, liu2022joint}, with a limited exploration in text generation tasks. More importantly, while these methods incorporate neighborhood information from knowledge graphs, they neglect two fundamental knowledge encoded in the topological pattern, proximity-based knowledge~\cite{rossi2020proximity, han2023towards, zhu2021node} (e.g., nodes that can be mutually reached via only few-hops walks) and role-based knowledge~\cite{rossi2014role, ahmed2018learning, ribeiro2017struc2vec} (i.e., nodes with similar local subgraph structures). For example, in Figure~\ref{fig-example}(a), when writing the abstract of a paper (node 1), having access to its cited papers in its related work section (nodes 2-4) could greatly enhance writing efficiency, because these referenced papers likely possess knowledge and narrative styles akin to the paper being authored. Likewise, in Figure~\ref{fig-example}(b), employees holding comparable positions in a company, like managers 2 and 3, typically share similar job responsibilities and analogous communication styles in their emails. Understanding the email content and job responsibilities of one manager can provide insights into the role and communication style of the other manager.

Given the overlook of the above two fundamental topological relations in current RAGs, we bridge the connection between topological and textual relations, and study their potential in enhancing RAGs for text generation. Our empirical analysis reveals the positive relation between the nodes' textual relations and their topological relations, which inspires us to explore the idea of augmenting text generation by incorporating topological relations. To validate the feasibility of this idea, we address two key questions: can the text generation be improved by incorporating additional texts? To answer the first question, we incrementally increase the contextual similarity to the text to be generated and observe a corresponding performance increase. Secondly, can textual relations between any pair of nodes be reflected by their topological relations? To answer the second question, we analyze the correlation between the textual relations between any two nodes and their proximity/role-based topological relations. Affirmative answers to the above two questions inspire us to propose a framework, Topology-aware Retrieval-augmented Generation (Topo-RAG), which augments text generation by retrieving relevant texts based on their proximity/role-based topological similarity. We pioneer the use of the task-oriented approach to verify the effectiveness of the Topo-RAG framework in text generation.

\begin{itemize}[leftmargin=*]
    \item \textbf{Bridging Topological and Textual Relations}: we discover the positive correlation between proximity/role-based topological relations of any pair of two nodes and their textual relations over nine datasets across four distinct domains, bridging the gap of node pairwise relations between topology space and text space.

    \item \textbf{Developing Topology-Informed RAG Framework}: we equip the RAG with topological awareness by retrieving texts from the graph database according to their proximity/role-based similarity to the target entity for which the text is being generated.

    \item \textbf{Comprehensive Empirical Analysis}: We construct a wide range of text-attributed graphs from diverse domains and conduct comprehensive experiments to verify the effectiveness of our framework. Moreover, we pioneer the use of node classification and link prediction to evaluate the quality of the generated texts.

\end{itemize}

\section{Preliminary}
\subsection{Motivative Examples}
\begin{figure}[t!]
    \centering
    \includegraphics[width=0.48\textwidth]{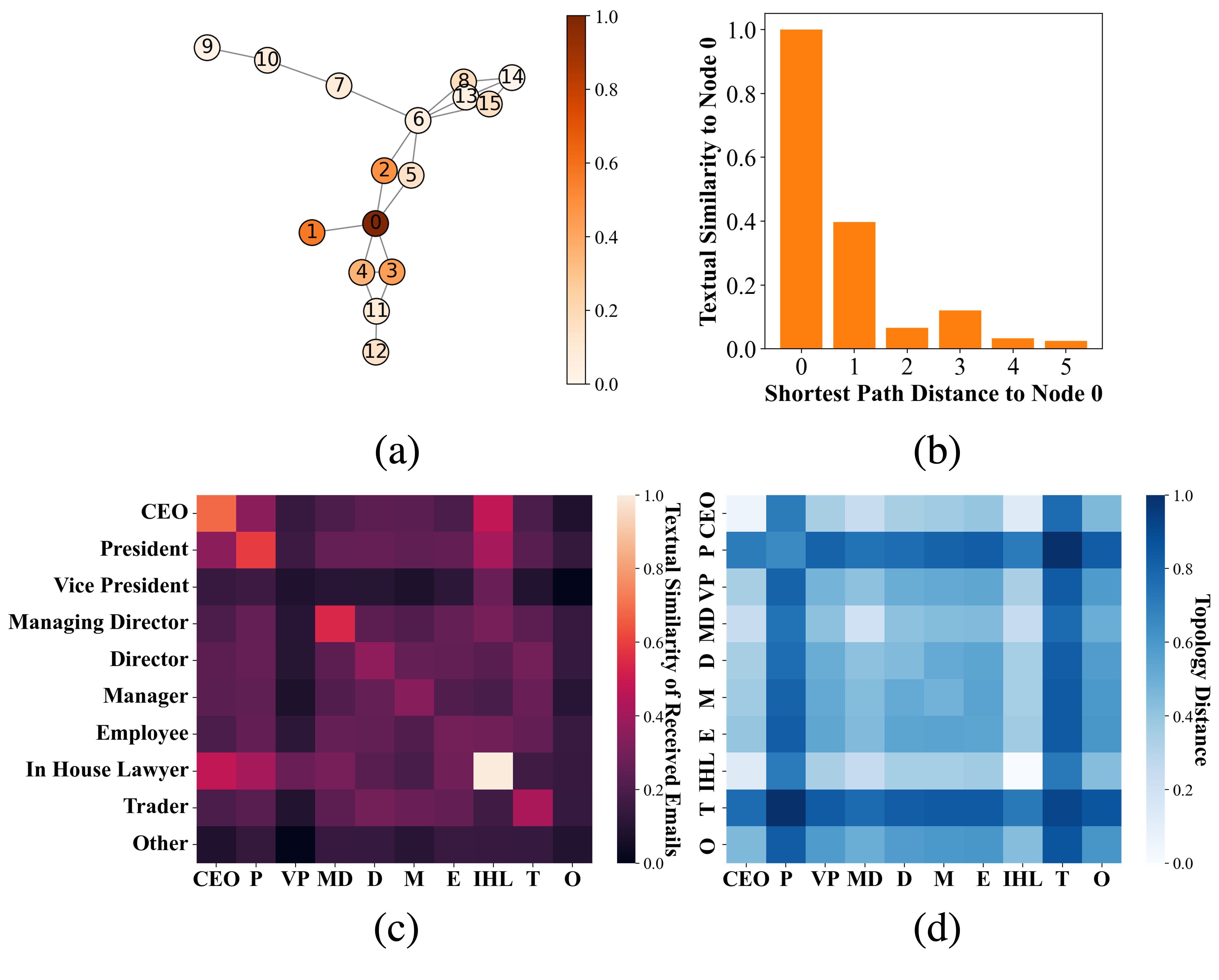}
    \vspace{-2ex}
    \caption{\textbf{(a)-(b)}: Cora citation network where nodes are papers and edges are reference relations. Papers that are topologically closer to paper 0 have higher textual similarity to its text; \textbf{(c)-(d)}: Eron-email network where nodes represent employees and edges denote their email communications. For each row in the two heatmaps, employees in diagonal entries share higher textual similarity and lower role-based topological distance than the ones belonging to off-diagonal entries in the same row. This indicates that employees with the same roles share a higher textual similarity and lower topological distance than employees with different roles.}
    \vspace{-2ex}
    \label{fig-motivation}
\end{figure}

To motivate the analysis of the correlation between the textual and topological relation, we take two examples, one analyzing the proximity-based relation (Figure~\ref{fig-motivation}(a)-(b)), and the other one analyzing the role-based relation (Figure~\ref{fig-motivation}(c)-(d)).

\subsubsection{Textual Relation with Proximity-based Topological Relation.}
In Figure~\ref{fig-motivation}(a), we extract a subgraph centering around node 0 from the citation network, Cora, where each node represents a paper with the abstract as its textual information, and each edge indicates a reference relation between the corresponding two paper nodes. Meanwhile, we obtain textual embedding of each node by feeding its abstract through sentence-transformer~\cite{reimers-2019-sentence-bert} and calculate the pairwise cosine similarity in Figure~\ref{fig-motivation}(b). Comparing Figure~\ref{fig-motivation}(a) and (b), we observe that as nodes become further away from node 0, their textual similarity to node 0 gradually decreases, indicating the textual similarity between any pair of nodes is correlated to their proximity-based topological distance.

\subsubsection{Textual Relation with Role-based Topological Relation.} We use the Eron-email dataset~\cite{creamer2022discovering} where we have access to emails sent from/received by employees with their job titles. We feed each email through the sentence-transformer~\cite{reimers-2019-sentence-bert} to obtain each email embedding. Then, we calculate the embedding of each employee by averaging the embeddings of their received emails. After that, we calculate pairwise textual similarity among employees and group them based on their job titles in Figure~\ref{fig-motivation}(c). For each row, we observe that the diagonal entry generally has higher textual similarity compared with other off-diagonal entries in the same row, indicating higher content similarity of emails received among employees possessing the same role in a company.\footnote{Similar observation is also found when analyzing their sent emails in Figure~\ref{fig-sent_emails} in Appendix~\ref{app-sent}.} Furthermore, we construct the graph with nodes being employees and edges being their email communications. We run GraphWave~\cite{donnat2018learning} to obtain employees' role-based embeddings and encode their local structural information. We calculate pairwise L2 distance among employees and group them based on their job titles. Similarly, we observe a generally lower role-based topological distance for employees possessing the same roles, indicating their local subgraph structures are aligned with their job titles in reality.

The above two observations motivate us to explore the potential of augmenting text generation by leveraging topology information. In the next, we introduce notations used throughout this paper and formulate the task of topology-aware text generation.

\subsection{Notations}
Let $\mathcal{S} = \{S_i\}_{i = 1}^{m + n}$ be the set of $m + n$ text sequences where $S_i$ represents the $i^{\text{th}}$ sequence. Assume we also have access to an additional graph connecting these textual sequences $G = (\mathcal{V}, \mathcal{E})$ where each node $v_i$ corresponds to a textual sequence $S_i$ and the edge $e_{ij}$ denotes the connection between node $v_i$ and $v_j$. Furthermore, the adjacency matrix of this graph is notated as $\mathbf{A}\in \mathbb{R}^{(m + n)\times (m + n)}$ where $\mathbf{A}_{ij} = 1$ if there is an edge $e_{ij}$ connecting $v_i$ and $v_j$, and $\mathbf{A}_{ij} = 0$ otherwise. Let $\mathcal{N}_{i}$ be the neighbors of node $v_i$. We formulate the topology-aware text generation in the following.

\subsection{Task Formulation}
Given a set of textual sequences $\mathcal{S} = \mathcal{S}^{\text{Full}} \cup \mathcal{S}^{\text{Partial}}$ where $\mathcal{S}^{\text{Full}} = \{S_i\}_{i = 1}^{n}$ comprises sequences with completely accessible texts, and $\mathcal{S}^{\text{Partial}} = \{S_i\}_{i = 1}^{m}$ consists of sequences with texts that are only partially observable. As the objective of many text-generation applications is to generate complete texts based on the first few pre-existing words/sentences, the "partially observed texts" in this paper refer to the initial words provided in a sequence. Let the partially observed text be $X_i$ for $i^{\text{th}}$ sequence and its unobserved counterpart be $Y_i$. We aim to leverage LLMs $\mathcal{F}$ to generate the sequence $\widehat{Y}_i$ utilizing the information from input $X_i$, other fully observed texts in $\mathcal{S}^{\text{Full}}$ and their topological relations in $G$, with the expectation to recover the ground-truth sequence $Y_i$:

\vspace{-2ex}
\begin{equation}\label{eq-colgen}
    \widehat{Y}_i = \mathcal{F}(\Omega(X_i, \mathcal{S}^{\text{Full}}, G)), ~~~~\forall i \in \{1, 2, ..., m\}.
\end{equation}

Comparing with solely relying on the partially observed text $X_i$ for the generation, i.e., $\widehat{Y}_i = \mathcal{F}(X_i)$, Eq.~\eqref{eq-colgen} leverages $\Omega$ to further retrieve the additional sequences from $\mathcal{S}^{\text{Full}}$ based on the topological knowledge in the graph structure $G$. The general hypothesis here is that for each pair of two textual sequences $S_i, S_j$ and their corresponding nodes $v_i, v_j$ in the graph, if 1) the text generation of LLMs can benefit from providing extra texts that are similar to the target text and 2) the textual similarity $\phi^{\text{Text}}_{ij} = \phi^{\text{Text}}(S_i, S_j)$ is correlated to their topological similarity $\phi^{\text{Topo}}_{ij} = \phi^{\text{Topo}}(v_i, v_j)$, then incorporating those topologically-similar sequences would benefit the generation of the current texts. Verifying the above hypothesis requires answering the following two questions:
\begin{itemize}[leftmargin=*]
    \item $\boldsymbol{Q_1}$: Would the text generation with a pre-trained large language model benefit from providing texts that have higher textual similarity to the current text to be generated?
    \item $\boldsymbol{Q_2}$: Is there any correlation between the textual similarity of two sequences and the topological similarity of their corresponding nodes?
\end{itemize}

Next, we address $\boldsymbol{Q}_1$ by empirically demonstrating the performance increase when providing additional texts with increasing textual similarity to the target text (Figure~\ref{fig-topo_rank} in Section~\ref{sec-when-benefit}). We address $\boldsymbol{Q}_2$ by formally introducing two types of topological similarity, proximity-based one and structure-based one, and analyzing their correlations to textual similarity (Figure~\ref{fig-proximity-analysis}/\ref{fig-structure-analysis} in Section~\ref{sec-topo-sim}). Successfully answering these two questions would motivate the proposal of our framework Topology-aware Retrieval-Augmented Generation (TopoRAG).

\section{Would text generation benefit from providing additional texts?}\label{sec-when-benefit}

\begin{figure}[t!]
    \centering
    \includegraphics[width=0.48\textwidth]{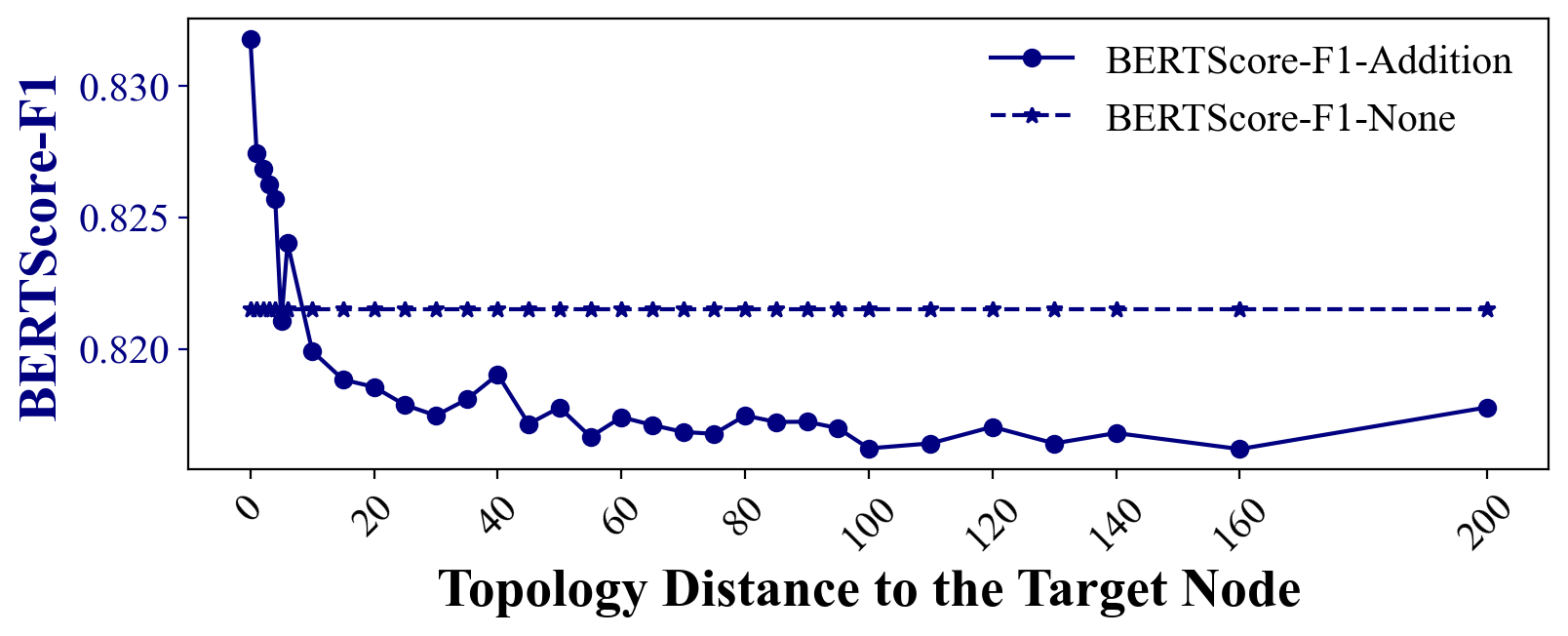}
    \caption{Comparing text generation between the scenario "Addition" where we include additional texts based on their proximity-based topological similarity to the target node and the scenario "None" where we include no additional texts but only based on its partially observed starting words.}
    \label{fig-topo_rank}
    \vspace{-4ex}
\end{figure}

To address the first research question $\boldsymbol{Q}_1$, we compute the topological similarity between the target node designated for text generation and all other nodes using their proximity-based topological embedding according to Eq~\eqref{eq-diffusion}. Then, we rank all other nodes based on the topological similarity to the target node, iteratively select three consecutive nodes from the ranked list (ranging from positions $k$ to $k + 3$), and incorporate their texts as supplementary context to enhance the text generation. Figure \ref{fig-topo_rank} shows the generation performance of paper abstracts on Cora as we sequentially increase the topological rank of the selected additional nodes.

Firstly, compared with solely based on its partially observed starting words (horizontal line), the text generation performance is better when including nodes that are among the Top 6. This finding highlights the benefits of incorporating additional texts in augmenting the current text generation. Secondly, by gradually increasing the rank of nodes we select, the performance gradually decreases according to BertScore-F1, demonstrating that the benefit of including additional texts is correlated to the similarity of those texts to the target one, answering $\boldsymbol{Q_1}$.

\section{Is topological similarity correlated to textual similarity?}\label{sec-topo-sim}
Assuming the topological similarity $\phi^{\text{Topo}}$ and the textual similarity $\phi^{\text{Text}}$ are two random variables where their specific realizations correspond to their values for a specific pair of nodes, then their Pearson correlation is computed as:
\begin{equation}
\tiny
    r = \frac{N^2\sum_{i, j = 1}^{N,N}\phi^{\text{Text}}_{ij}\phi^{\text{Topo}}_{ij} - \sum_{i,j = 1}^{N, N} \phi^{\text{Text}}_{ij}\sum_{i, j = 1}^{N, N}\phi^{\text{Topo}}_{ij}}{\sqrt{[N^{2}\sum_{i,j=1}^{N, N}{(\phi^{\text{Text}}_{ij})}^2 - (\sum_{i,j=1}^{N, N} \phi_{ij}^{\text{Text}})^2][N^{2}\sum_{i,j=1}^{N, N}{(\phi^{\text{Topo}}_{ij})}^2 - (\sum_{i,j=1}^{N, N} \phi_{ij}^{\text{Topo}})^2]}},
\end{equation}
where $\phi_{ij}^{\text{Text}} = \phi^{\text{Text}}(S_i, S_j)$ is defined as the semantic similarity and it is computed as the cosine similarity of textual embeddings from sentence-transformer. $\phi_{ij}^{\text{Topo}} = \phi^{\text{Topo}}(v_i, v_j)$ defines the topological similarity between two nodes $v_i, v_j$. Following previous works~\cite{ribeiro2017struc2vec, donnat2018learning}, we measure this topological similarity between two nodes by the similarity of their topological embeddings. In the next, we explore two types of node topological embeddings: proximity-based and role-based ones.

\subsection{Proximity-based Topological Similarity}\label{sec-cor-proxi}
Proximity-based topological similarity quantifies the similarity between two nodes via their topological distance in the graph. The general intuition here is that two topologically close nodes usually have a higher textual similarity.

Conventional ways of characterizing the topological distance between two nodes in a graph include the shortest path distance and the shallow embeddings (e.g., random walk, personalized page rank and diffusion~\cite{grover2016node2vec, gasteiger2019diffusion}). In this work, we choose the diffusion-based one due to its thorough consideration of all potential paths in contributing to the proximity between any two nodes. Given a degree-normalized adjacency matrix $\widehat{\mathbf{A}} = \mathbf{D}^{-1}\mathbf{A}$ and an identity matrix $\mathbf{I}\in \{0, 1\}^{N\times N}$ uniquely identifying each node, we perform diffusion to obtain the propagated node embeddings as:
\begin{equation}\label{eq-diffusion}
    \mathbf{P} = \sum_{k = 1}^{K}\alpha_{k}\widehat{\mathbf{A}}^k\mathbf{I},
\end{equation}
where $\mathbf{P}_{ij}$ quantifies the influence of node $v_j$ on node $v_i$ considering paths of lengths varying from $1$ to $K$ since two nodes that are topologically close to each other should receive similar influence from all other nodes in this graph. However, directly performing this diffusion requires $\mathcal{O}(KN^3)$ for time and $\mathcal{O}(N^2)$ for space complexity. To make this computation scalable~\cite{chen2019fast}, we further project the unique node label matrix $\mathbf{I}$ via random gaussian projection by replacing $\mathbf{I}\in\mathbb{R}^{N\times N}$ with $\mathbf{R} \in \mathbb{R}^{N\times d}\sim \mathcal{N}(\mathbf{0}^{d}, \boldsymbol{\Sigma}^{d})$ with $d << N$, which effectively reduces the time/space complexity to $\mathcal{O}(KN^2d)$/$\mathcal{O}(Nd)$. Then the topological similarity between two nodes is computed as the cosine similarity between their corresponding propagated node embeddings, i.e., $\phi^{\text{Topo}}_{i, j} = \text{cos}(\mathbf{P}_i, \mathbf{P}_j)$.

\begin{figure}[t!]
    \centering
    \includegraphics[width=0.48\textwidth]{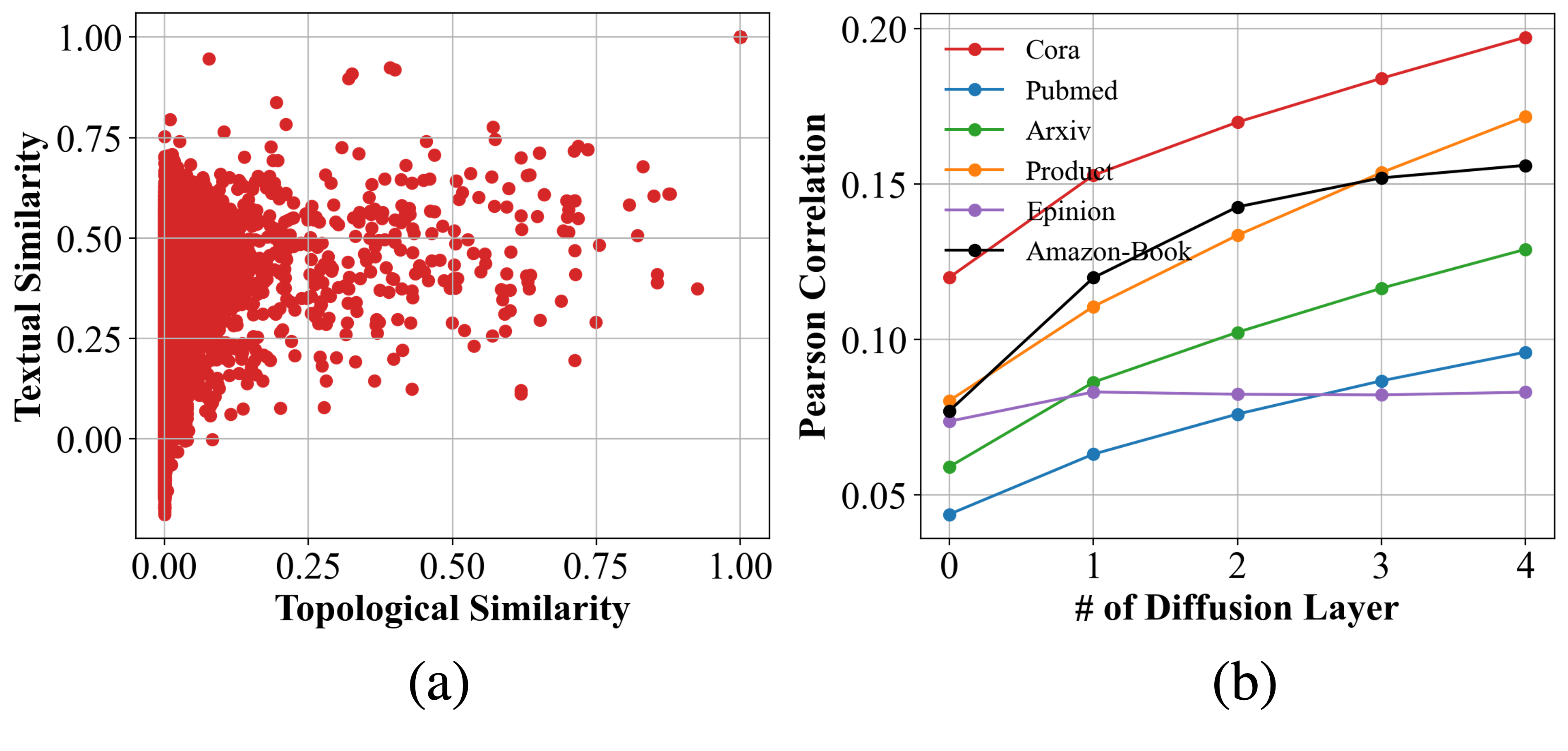}
    
    \caption{Correlation between Proximity-based Topological Similarity and Textual Similarity. \textbf{(a)}: in Cora, as the topological distance between two paper nodes decreases, their textual similarity increases. \textbf{(b)}: the Pearson correlations across different datasets are all positive and increase as the number of diffusion layer $k$ in Eq.~\eqref{eq-diffusion} increases.}
    \label{fig-proximity-analysis}
\end{figure}

To verify the correlation between the proximity-based topological similarity and textual similarity, we conduct correlation analysis on datasets in Figure~\ref{fig-proximity-analysis}. On Cora dataset, we run diffusion according to Eq.~\eqref{eq-diffusion}, obtain node embeddings $\mathbf{P}$, calculate the pairwise topological similarity, and visualize it along with the pairwise textual similarity in Figure~\ref{fig-proximity-analysis}(a)~\footnote{Similar analysis on other datasets are included in Figure~\ref{fig-extra-corr} in Appendix~\ref{app-cor}}. We can see the pairwise textual similarity increases as the pairwise topological similarity increases. Moreover, we calculate the Pearson correlation across different datasets at different diffusion layers $k$ in Eq.~\eqref{eq-diffusion} in Figure~\ref{fig-proximity-analysis}(b). In all six datasets, the correlation is positive and increases as the diffusion layer increases as the higher diffusion layer considers the higher-order paths in quantifying the topological proximity, which becomes more aligned with their textual similarity. Different from Cora/Pubmed/Arxiv/Product/Amazon-Book, the correlation on the Epinion dataset does not increase since reviews posted by the same person may not necessarily be similar if the products are different.

\begin{figure}[t!]
    \centering
    \includegraphics[width=0.48\textwidth]{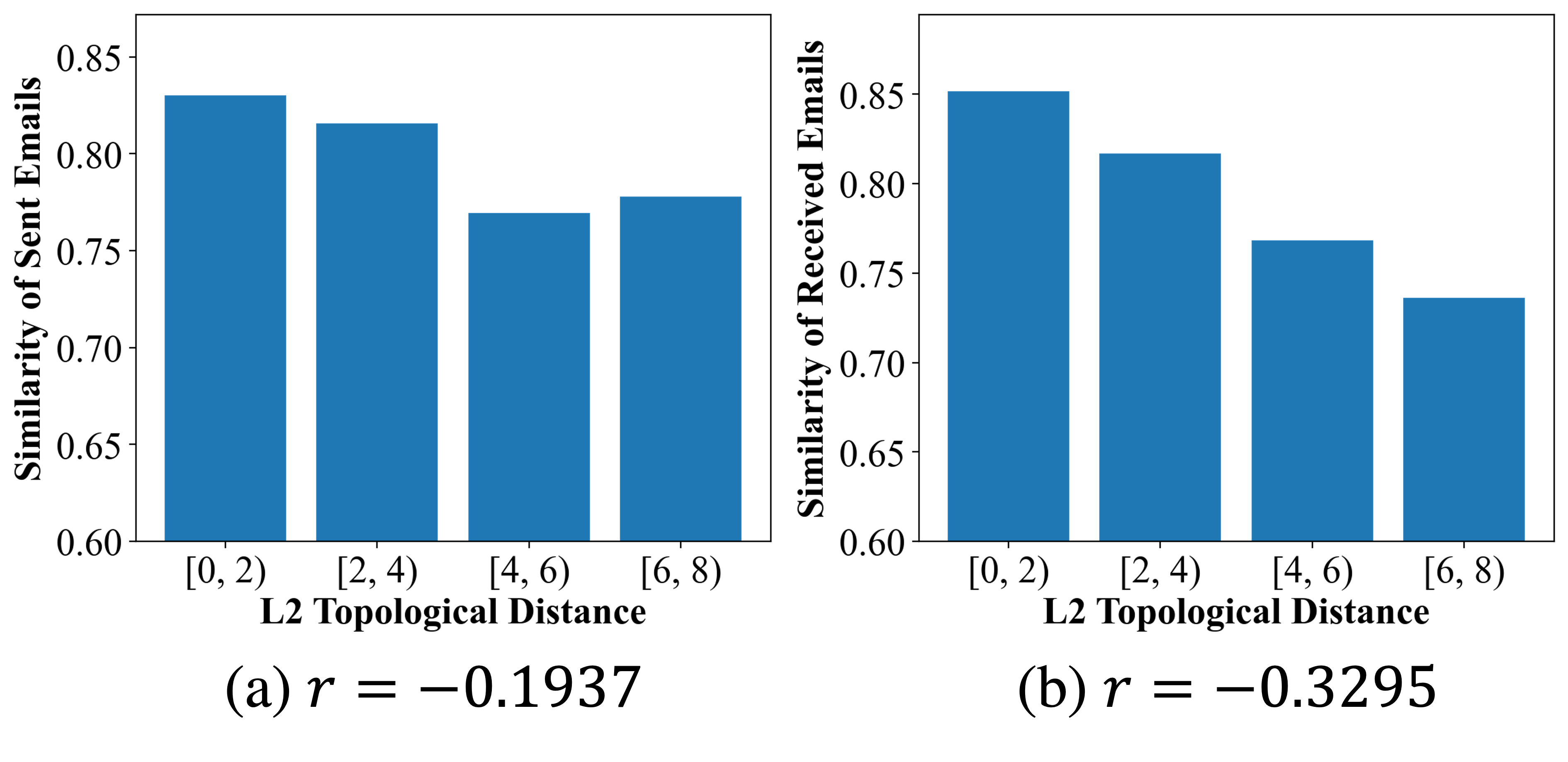}
    
    \caption{Correlation between Role-based Topological Distance and Textual Similarity on Eron-Email Dataset where role-based topological distance is calculated based on the L2-distance between embeddings of two employees obtained from GraphWave while textual similarity is calculated as the average cosine similarity of textual embeddings of either the sent~(a) or received emails~(b) between two employees.}
    \label{fig-structure-analysis}
\end{figure}

\subsection{Role-based Topological Similarity}\label{sec-cor-structure}
Unlike proximity-based topological similarity which considers nodes residing closely in one network to be similar, role-based topological similarity focuses on identifying nodes with topologically similar neighborhoods~\cite{ribeiro2017struc2vec}. Intuitively, nodes with similar local structures perform similar functions in the network and hence possess similar textual information in some aspects, such as the employees' tone should be different from the vice presidents' tone in a company~\cite{henderson2012rolx}. Following this intuition, we employ one of the most representative embedding methods, GraphWave, that encodes the node local structure information in the next.

Following GraphWave~\cite{donnat2018learning}, the spectral graph wavelet $\Psi_a$ is calculated as:
\begin{equation}
\Psi_a = \mathbf{U} \text{Diag}(g_s(\lambda_1), \dots, g_s(\lambda_N))\mathbf{U}^{\top} \mathds{1}_a,
\end{equation}
where $\mathbf{L} = \mathbf{D} - \mathbf{A} = \mathbf{U} \mathbf{\Lambda} \mathbf{U}^{\top}$ and $\lambda_1 \le \lambda_2 \le ... \le \lambda_N (\mathbf{\Lambda} = \text{Diag}(\lambda_1, \lambda_2, ..., \lambda_N))$ is the eigenvalues of $\mathbf{L}$) with $g_s(\lambda) = e^{-\lambda s}$ being the kernel filter. $\mathds{1}_a$ is the one-hot vector for node $v_a$. $\Psi_a$ denotes the spectral graph wavelet for the heat kernel centered at node $v_a$ and more specifically, its $b^{\text{th}}$-entry $\Psi_{ba}$ denotes the amount of energy that node $v_a$ has received from node $v_b$. Furthermore, GraphWave treats the spectral graph wavelet of node $v_a$, i.e., $\Psi_a$, as a probability distribution and characterizes it via the empirical characteristic function with $i$ being the imaginary number here:
\begin{equation}
    \phi_{\Psi_a}(t) = \mathbb{E}_{v_b \sim \mathcal{V}}(e^{it\Psi_{ab}}),
\end{equation}
Eventually, structural embedding $\mathbf{P}_a$ of node $v_a$ is obtained by sampling the 2-dimensional characteristics function at d evenly spaced points $t_1, t_2, ..., t_d$ and concatenating them as:
\begin{equation}\label{eq-role}
    \mathbf{P}_a = ||_{t_1, t_2, ..., t_d}[\text{Re}(\phi_{\Psi_a}(t)), \text{Im}(\phi_{\Psi_a}(t))].
\end{equation}
Then the topological similarity between two nodes is computed as the L2-distance between their corresponding structural embeddings, i.e., $\phi_{a,b}^{\text{Topo}} = \frac{1}{||\mathbf{P}_a - \mathbf{P}_b||_2}$.

To verify the correlation between the role-based topological similarity and textual similarity, we conduct correlation analysis on the Eron-Email dataset in Figure~\ref{fig-structure-analysis}.  Following the experimental setting for Figure~\ref{fig-motivation}(c)-(d), we group each pair of employees based on their L2 topological distance and compute average textual similarity using their sent/received emails in Figure~\ref{fig-structure-analysis}(a)/(b). We can see the negative correlation (-0.1937/-0.3295) between the role-based topological distance between two employees and the similarity of their sent/received emails.

To sum up, the positive responses to the preceding questions confirm two key insights. Firstly by the answer to $\boldsymbol{Q}_1$, LLMs significantly benefit from incorporating additional texts during text generation. The closer the resemblance of these additional texts to the generated one, the greater the enhancement is in the output. Secondly, by the answer to $\boldsymbol{Q}_2$, there is a positive correlation between textual and topological similarity. Drawing from these findings, we propose a novel framework using topological similarity to guide the retrieval of additional texts, thereby augmenting the quality of the generated text, i.e., Topology-aware Retrieval-Augmented Generation (Topo-RAG), which is introduced next.
\section{Framework of Topo-RAG}\label{sec-TopoRAG}
Following Eq.~\eqref{eq-colgen}, the retriever works by retrieving top-K nodes according to their topological similarity to the target text as follows:
\begin{equation}\label{eq-col-retrieval}
     \mathcal{T}_i = \Omega^{\text{Retriever}}(X_i, \mathcal{S}^{\text{Full}}, G) = \underset{v_j \in \mathcal{V}}{\argmax}^K \phi^{\text{Topo}}(v_i, v_j),
\end{equation}

After that, we formulate the texts of nodes in $\mathcal{T}_i$ along with the partially observed text $X_i$ of target node $v_i$ into the prompt triggering LLM for text generation. Compared with conventional retrievals that only consider textual relations, the Topo-RAG framework considers topological relations.  To implement this, it's necessary to pre-calculate the topological relations between every pair of nodes, a process requiring $\mathcal{O}(|\mathcal{V}|^2)$ in both space and time complexity. To manage space constraints, we apply a top-k thresholding, reducing the space requirement significantly to $\mathcal{O}(|\mathcal{V}|K)$. These computations are done in advance during an offline phase. Consequently, when a text generation request for a specific target node is made, we can retrieve the necessary information instantaneously from the pre-computed dictionary, maintaining a time complexity of $\mathcal{O}(1)$. Note that prompting LLMs with the retrieved nodes of very long texts could exceed the input limits. Since this is a common issue for any RAG framework, one can equip our Topo-RAG framework with the existing strategies~\cite{jin2024llm, wang2024augmenting} that handle this long-context issue. Here, we exclude instances where the context limit is exceeded. The length distribution analysis in Figure~\ref{fig-sent_emails}(b) reveals that the textual length of most nodes remains within acceptable limits, ensuring that our results and insights are still applicable to the original dataset even after the exclusion.
\section{Experiments}\label{sec-experiments}

\begin{table}[h!]
\footnotesize
\setlength{\extrarowheight}{2pt}
\setlength\tabcolsep{3.5pt}
\caption{Statistics of Datasets. $\mathcal{S}^{\text{Full}}$ denote nodes with fully available textual information while $\mathcal{S}^{\text{Partial}}$ denote nodes with partially observable text.}
\begin{tabular}{llcccc}
\hline
\textbf{Domain} & \textbf{Dataset} & \textbf{\# Nodes} & \textbf{\# Edges} & \makecell{\textbf{\# Instances}\\ ($\mathcal{S}^{\text{Full}}/\mathcal{S}^{\text{Partial}}$)}  & \textbf{Splitting}\\
\hline
\multirow{3}{*}{Citation} & Cora~\cite{yang2016revisiting, chen2023exploring} & 2,708 & 5,429 & 2,522/186 & Random\\
 & Pubmed~\cite{yang2016revisiting, chen2023exploring} & 19,717 & 44,335 & 17,786/1,931 & Random \\
 & Arxiv~\cite{hu2020open, chen2023exploring} & 16,316 & 53,519 & 14,791/1,525 & Time \\
\hline
\multirow{5}{*}{E-commerce} & Product~\cite{hu2020open, chen2023exploring} & 16,475 & 60,015 & 15,790/685 & Random \\
 & Book~\cite{ni2019justifying} & 7,252 & 203,438 & 6,526/726 & Time \\
 & Epinion~\cite{cai2017spmc, zhao2015improving} & 4,976 & 15,613 & 4,477/499 & Time \\
& Music~\cite{ni2019justifying} & 10,341 & 447,250 & 9,306/1,035 & Time \\
& Pantry~\cite{ni2019justifying} & 4,650 & 43,970 & 4,184/466 & Time \\
\hline
Social 
 & Eron-Email~\cite{shetty2004enron} & 18,055 & 123,208 & 182,265/46,990 & Random \\
\hline
\end{tabular}
\vspace{-2ex}
\end{table}

\subsection{Experimental Settings}
\subsubsection{Datasets}
Although previous works~\cite{huo2019graph, koncel2019text} borrow knowledge graphs to enhance text generation, the improvement mainly comes from the complex logical pattern encoded in the knowledge graph rather than proximity/structure topological patterns discussed in this paper. Therefore, we collect additional datasets to demonstrate the effectiveness of considering these two patterns in text generation, the details of which are discussed next:

\begin{itemize}[leftmargin=*]
    \item \textbf{Cora, Pubmed, Arxiv}~\cite{yang2016revisiting, chen2023exploring, hu2020open}: Citation networks where nodes represent papers with abstracts as textual information, and edges signify reference relations. We divide nodes into fully-observed/partially-observed sets in a 90\%/10\% ratio and further remove nodes whose abstracts are less than 100 words. Then, we create the induced subgraph and remove edges connecting two nodes in the testing set to avoid information leakage. For the larger Arxiv network, due to resource constraints, we randomly select 2\% nodes as seeds and apply the GraphSAGE sampling with the number of neighbors [2, 2] across two layers. Since Arxiv provides the publication time of each paper~\cite{hu2020open}, we use the same preprocessing as Cora/Pubmed but follow the chronological order, imitating the real scenarios where users are writing papers with references to historical papers.

    \item \textbf{Book, Epinion, Music, Pantry}~\cite{hu2020open, chen2023exploring, ni2019justifying, cai2017spmc, zhao2015improving}: In digital e-commerce networks, each node represents a review and two reviews have an edge if they are either written by the same customer or posted on the same product. Considering the vast scale of the Book dataset, which encompasses 27,161,262 reviews, we only take the latest 1\% reviews to construct the graph. We exclude any review whose length falls below 100 characters. We adhere to the same data splitting ratio used in Cora/Pubmed/Arxiv following the chronological order when the review was generated.

    \item \textbf{Products}~\cite{hu2020open}: Amazon product co-purchasing network where nodes represent products sold in Amazon and edges between two products indicate the co-purchase behaviors. Following Arxiv, we randomly select 0.1\% among all product nodes as seeds and apply a GraphSAGE-based neighborhood sampling, with the number of neighbors [2, 2] across two layers. Different from~\cite{hu2020open} using the sales ranking to split nodes, we randomly select 90\%/10\% nodes into fully-observed/partially-observed sets.

    \item \textbf{Eron-Email}~\cite{donnat2018learning}: In email communication networks, each node represents an employee in the company with his/her textual information being the historical written/received emails. We preprocess the original emails and extract the sender/receiver/text information of each email following script here\footnote{\href{https://github.com/mihir-m-gandhi/Enron-Email-Analysis/tree/main}{https://github.com/mihir-m-gandhi/Enron-Email-Analysis/tree/main}}.
\end{itemize}

\subsubsection{Baselines}
For baselines, as no previous works have considered the proximity/role-based relations in RAG, we design baselines by equipping LLMs, GPT3 and GPT3.5 for text generation, with the following RAG strategies: 
\begin{itemize}[leftmargin=*]
    \item \textbf{None}: we do not retrieve any additional texts but completely rely on the partially observed starting words.

    \item \textbf{Random (RD)}: we randomly retrieve $K$ texts from the graph-structured knowledge base.

    \item \textbf{Text}: we calculate the semantic similarity between the partially observed texts in the target sequence and all other texts. Then we select the Top-$K$ ones according to their semantic similarity.

    \item \textbf{Topo}: we calculate the topological similarity according to embeddings from Eq.~\eqref{eq-diffusion}/\eqref{eq-role} between the node of the target sequence and all other nodes. Then we rank them and select the top-$K$ ones. This one is essentially our Topo-RAG framework.
\end{itemize}

\subsubsection{Evaluation Tasks.} To evaluate the quality of our generated text, we compare it with ground-truth text following established methodologies~\cite{li2023teach, salemi2023lamp}. We only focus on comparing the generated content, without considering any initially observed words. In addition, we introduce a task-oriented evaluation to assess the quality of the generated texts. Specifically, we adopt two graph-based tasks, node classification and link prediction. In these two tasks, textual features of certain nodes are assumed to be reconstructed using various baselines including our Topo-RAG. With the reconstructed textual features of nodes after text generation, we then train graph machine learning models and evaluate their performance. We conduct the evaluation using GCN~\cite{kipf2016semi}, SAGE~\cite{hamilton2017inductive}, and MLP to exclude any model-induced bias during evaluation.

\subsubsection{Evaluation Metrics}
Following conventional works~\cite{li2023teach, salemi2023lamp}, we use the BLEU-4/ROUGE-L/Bert-F1 score as the evaluation metrics to conduct a comprehensive analysis of the generated texts. In addition, we take the initiative to use the task-oriented evaluation, which quantifies the quality of generated texts based on whether they can fulfill purposes of downstream tasks, e.g., node classification and link prediction. For node classification, we report the average accuracy of testing nodes (in our setting, the testing nodes are assumed to be the ones with features to be reconstructed). For link prediction, we report the average Hits@100 following~\cite{hu2020open} of randomly selected edges.

\subsubsection{Parameter Settings}
For text generation, the number of retrieved texts and partially observed starting words are both set as 3. Moreover, we set the number of generated words to be 150, 250, 200, 150, 300, 500, 250, 200, 300 according to the average length of texts in Cora, Pubmed, Arxiv, Product, Book, Epinion, Music, Pantry, Eron-Email. For all datasets except Eron-Email, each node is only associated with one text sequence, hence we could directly calculate textual/topological similarity metric and retrieve the Top-3 accordingly. For Eron-Email where each node/employee possesses many texts/emails, we first query the sender and receiver of the target email to be generated and then collect emails from the two employees with the highest topological similarity to that sender and receiver. Furthermore, we select the Top-3 emails from those collected emails based on their textual similarity to the partially observed target text. For most of the hyperparameters used for evaluation with node classification and link prediction, we follow the same setting as \cite{wang2021tree} and \cite{zhao2022learning}. In node classification, the hyperparameters are: training epoch is 1000, learning rate 0.01, weight decay 0.0005, early stopping 100, 2 layer graph convolution layer/MLP, dropout 0.5, number of hidden layers 64. In link prediction, the hyperparameters are: encoder learning rate 0.001, predictor learning rate 0.001, number of hidden layers 256, and dropout 0.

\vspace{-2ex}
\subsection{Performance Comparison}\label{sec-perform}

\begin{table*}[t!]
\small
\caption{Performance comparison of TopoRAG with baselines. The best results are in \textbf{bold}. BLEU is BLEU-4, ROUGE is ROUGE-L. Our TopoRAG almost achieves the best performance across all baselines on all datasets. "Average" is computed by averaging each metric across 9 datasets. "Boost" is computed by the relative performance gain from the second-to-best "Text" to the best "TopoRAG".}
\vspace{-1ex}
\setlength{\extrarowheight}{1.8pt}
\setlength\tabcolsep{3pt}
\begin{tabular}{l|l|ccc|ccc|ccc|ccc|ccc}
\Xhline{2\arrayrulewidth}
\multirow{2}{*}{\textbf{LLM}} & \multirow{2}{*}{\textbf{Retriever}} & \multicolumn{3}{c|}{\textbf{Cora}} & \multicolumn{3}{c|}{\textbf{Pubmed}} & \multicolumn{3}{c|}{\textbf{Arxiv}} & \multicolumn{3}{c|}{\textbf{Product}} & \multicolumn{3}{c}{\textbf{Book}} \\
 &  & BLEU & ROUGE & Bert-F1 & BLEU & ROUGE & Bert-F1 & BLEU & ROUGE & Bert-F1 & BLEU & ROUGE & Bert-F1 & BLEU & ROUGE & Bert-F1 \\
\Xhline{2\arrayrulewidth}
\multirow{5}{*}{{\makecell{\textbf{GPT}\\\textbf{3.5}}}} & \textbf{None} & 1.46 & 15.90 & 82.15 & 1.52 & 14.63 & 80.28 & 1.22 & 15.25 & 81.87 & 1.57 & 15.09 & 81.98 & 1.05 & 14.75 & 80.98\\
 & \textbf{RD} & 1.78 & 16.42 & 83.10 & 2.38 & 15.83 & 81.32 & 2.64 & 16.22 & 83.03 & 1.65 & 14.90 & 82.09 & 1.34 & 14.77 & 82.30\\
 & \textbf{Text} & 1.77 & 16.43 & 83.05 & 2.27 & 15.49 & 81.37 & 2.23 & 15.89 & 82.96 & 2.44 & 15.50 & 82.16 & 1.77 & 15.30 & 82.53\\
 \cline{2-17}
 & \textbf{TopoRAG} & \textbf{3.49} & \textbf{17.58} & \textbf{83.86} & \textbf{3.97} & \textbf{17.54} & \textbf{82.97} & \textbf{3.66} & \textbf{17.49} & \textbf{84.10} & \textbf{3.65} & \textbf{16.85} & \textbf{83.17} & \textbf{2.55} & \textbf{16.14} & \textbf{83.15}\\
 & \textbf{Boost} & 97.18\% & 7.00\% & 0.98\% & 74.89\% & 13.23\% & 1.97\% & 64.13\% & 10.07\% & 1.37\% & 49.59\% & 8.71\% & 1.23\% & 44.07\% & 5.49\% & 0.75\%\\
 \hline
\multirow{5}{*}{{\makecell{\textbf{GPT}\\\textbf{3}}}} & \textbf{None} & 1.16 & 15.40 & 81.07 & 1.19 & 14.10 & 79.42 & 0.88 & 14.47 & 80.53 & 1.90 & 14.71 & 80.85 & 0.75 & 14.20 & 80.03\\
 & \textbf{RD} & 1.72 & 16.45 & 82.79 & 2.48 & 15.88 & 81.11 & 2.50 & 16.44 & 82.90 & 1.48 & 14.50 & 81.16 & 1.00 & 15.10 & 82.18\\
 & \textbf{Text} & 2.21 & 16.50 & 82.85 & 2.32 & 15.69 & 81.24 & 2.17 & 15.82 & 82.60 & 2.93 & 15.54 & 81.24 & 1.15 & 15.30 & 81.83\\
 \cline{2-17}
 & \textbf{TopoRAG} & \textbf{4.19} & \textbf{17.58} & \textbf{83.69} & \textbf{4.20} & \textbf{17.67} & \textbf{82.89} & \textbf{3.65} & \textbf{17.48} & \textbf{83.95} & \textbf{4.91} & \textbf{17.33} & \textbf{82.88} & \textbf{1.88} & \textbf{15.80} & \textbf{82.78}\\
 & \textbf{Boost} & 89.59\% & 6.55\% & 1.01\% & 81.03\% & 12.62\% & 2.03\% & 68.20\% & 10.49\% & 1.63\% & 67.58\% & 11.52\% & 2.02\% & 63.48\% & 3.27\% & 1.16\%\\
 \Xhline{2\arrayrulewidth}
\end{tabular}
\hspace{0ex}
\vspace{0.5ex}
\begin{tabular}{l|l|ccc|ccc|ccc|ccc|ccc}
\Xhline{2\arrayrulewidth}
\multirow{2}{*}{\textbf{LLM}} & \multirow{2}{*}{\textbf{Retriever}} & \multicolumn{3}{c|}{\textbf{Epinion}} & \multicolumn{3}{c|}{\textbf{Pantry}} & \multicolumn{3}{c|}{\textbf{Eron-Email}} & \multicolumn{3}{c|}{\textbf{Music}} & \multicolumn{3}{c}{\textbf{Average}} \\
 &  & BLEU & ROUGE & Bert-F1 & BLEU & ROUGE & Bert-F1 & BLEU & ROUGE & Bert-F1 & BLEU & ROUGE & Bert-F1 & BLEU & ROUGE & Bert-F1 \\
\Xhline{2\arrayrulewidth}
\multirow{5}{*}{\makecell{\textbf{GPT}\\\textbf{3.5}}} & \textbf{None}  & 0.47 & 10.46 & 78.80 & 0.99 & 14.63 & 81.23 & 2.40 & 9.18 & 79.27 & 1.33 & 14.07 & 80.37 & 1.33 & 13.77 & 80.77\\
 & \textbf{RD} & 0.62 & 10.89 & 80.34 & 1.25 & 15.16 & 82.69 & 2.06 & 9.71 & 80.33 & 1.85 & 14.17 & 81.70 & 1.73 & 14.23 & 81.88\\
 & \textbf{Text} & 0.59 & 10.75 & 80.29 & 1.79 & 15.40 & 82.68 & 3.09 & 10.60 & 79.92 &  1.24 & 14.20 & 81.63 & 1.91 & 14.40 & 81.84 \\
 \cline{2-17}
 & \textbf{TopoRAG} & \textbf{1.00} & \textbf{11.22} & \textbf{80.75} & \textbf{2.03} & \textbf{15.65} & \textbf{83.00} & \textbf{3.98} & \textbf{11.83} & \textbf{80.70} & \textbf{3.49} & \textbf{15.08} & \textbf{82.13} & \textbf{3.09} & \textbf{15.49} & \textbf{82.65} \\
 & \textbf{Boost} & 69.49\% & 4.37\% & 0.57\% & 13.41\% & 1.62\% & 0.39\% & 28.80\% & 11.60\% & 0.98\% & 181.5\% & 6.20\% & 0.61\% & 61.78\% & 7.57\% & 0.99\%\\
 \hline
\multirow{5}{*}{\makecell{\textbf{GPT}\\\textbf{3}}} & \textbf{None} & 0.34 & 10.68 & 78.42 & 0.72 & 13.57 & 80.03 & 1.65 & 7.62 & 78.42 & 1.19 & 13.70 & 79.43 & 1.09 & 13.16 & 79.80 \\
 & \textbf{RD} & 0.48 & 10.86 & 80.16 & 1.04 & 14.98 & 82.49 & 3.30 & 10.55 & 79.49 & 1.17 & 14.25 & 81.53 & 1.69 & 14.33 & 81.53 \\
 & \textbf{Text} & 0.38 & 10.85 & 79.54 & 1.64 & 15.28 & 82.72 & 3.65 & \textbf{11.14} & 79.79 & 4.03 & 15.54 & 81.81 & 2.28 & 14.63 & 81.51\\
 \cline{2-17}
 & \textbf{TopoRAG} & \textbf{1.14} & \textbf{11.68} & \textbf{80.79} & \textbf{2.02} & \textbf{15.51} & \textbf{82.86} & \textbf{3.97} & 10.69 & \textbf{80.38} & \textbf{5.13} & \textbf{16.10} & \textbf{82.18}  & \textbf{3.45} & \textbf{15.54} & \textbf{82.49} \\
 & \textbf{Boost} & 200.0\% & 7.65\% & 1.57\% & 23.17\% & 1.51\% & 0.17\% & 8.77\% & -4.04\% & 0.74\% & 27.3\% & 3.6\% & 0.45\% & 51.32\% & 6.22\% & 1.2\%\\
 \Xhline{2\arrayrulewidth}
\end{tabular}
\label{tab-main}
\end{table*}

\subsubsection{Traditional Evaluation}
Here we compare the text generation capabilities of GPT-3.5 and GPT-3 enhanced with our proposed Topo-RAG and other baseline methods. Due to resource constraints, we only randomly select 500 nodes along with their partially observed textual sequences to complete and report the average performance in Table~\ref{tab-main}. Overall, the proposed TopoRAG framework achieves the highest text generation performance with a significantly large margin, as shown by "Average", underscoring the benefits of integrating topological knowledge for retrieving additional contextual information in text generation. The second to best baseline is "Text" because it utilizes textual similarity to directly query relevant context, which augments the text generation to some extent. Interestingly, we also find that including random texts in the generation process, i.e., "RD", significantly improves performance compared to "None" which includes no additional texts at all. This improvement likely arises because each text within the same text-attributed network is generally related to the same domain, e.g., all texts in Cora are papers and all texts in Epinion are reviews. Hence, incorporating additional texts can provide insights into potential writing styles and usage of domain-specific terminology, which benefits the text generation of the target node. Comparing performance boosts across different evaluation metrics reveals that improvements under BLEU-4 often result in relatively larger gains compared to those under ROUGE-L and BERT-F1. This is because BLEU-4 emphasizes exact matches of words and phrases in the generated text against the reference, making it particularly sensitive to precise word matching. Conversely, ROUGE-L measures the overlap of n-grams between generated and reference texts, regardless of their order or precise phrasing, rendering it less sensitive than BLEU. BERT-F1, which evaluates semantic meaning, is even less sensitive to exact word matches, focusing more on the contextual alignment of the content. Overall, this suggests that incorporating additional context into text generation is more beneficial for achieving similar terminology or writing style rather than for capturing general meaning.

\begin{table}[t!]
\small
\caption{Task-oriented Evaluation by comparing the node classification and link prediction performance of different baselines. The best results are in \textbf{bold}. NC - Node Classification; LP - Link Prediction.}
\renewcommand{\arraystretch}{0.8}
\setlength{\extrarowheight}{2pt}
\setlength\tabcolsep{3pt}
\begin{tabular}{l|l|cccc}
\Xhline{2\arrayrulewidth}
\multirow{2}{*}{\textbf{Model}} & \multirow{2}{*}{\textbf{Retriever}} & \multicolumn{2}{c}{\textbf{Cora}} & \multicolumn{2}{c}{\textbf{Pubmed}} \\
 &  & NC & LP & NC & LP \\ 
 \Xhline{2\arrayrulewidth}
\multirow{4}{*}{\textbf{GCN}} & None & 70.48$\pm$0.52 & 76.31$\pm$0.81 & 72.88$\pm$0.15 & 79.16$\pm$0.54 \\
 & RD & 70.68$\pm$1.05 & 75.41$\pm$0.81 & 72.98$\pm$0.44 & 78.78$\pm$0.86 \\
 & Text & 71.09$\pm$0.50 & 77.15$\pm$0.49 & 72.30$\pm$1.10 & 79.50$\pm$0.86 \\
 \cline{2-6}
 & TopoRAG & \textbf{75.49$\pm$0.26} & \textbf{89.12$\pm$0.49} & \textbf{77.80$\pm$0.53} & \textbf{81.33$\pm$0.40} \\ 
 \hline
\multirow{4}{*}{\textbf{SAGE}} & None & 60.75$\pm$1.34 & 80.44$\pm$0.98 & 64.24$\pm$3.47 & 79.77$\pm$0.62 \\
 & RD & 57.20$\pm$1.69 & 78.61$\pm$0.58 & 64.76$\pm$2.43 & 80.00$\pm$0.72 \\
 & Text & 57.07$\pm$2.95 & 82.40$\pm$0.61 & 64.18$\pm$2.28 & 80.48$\pm$0.15 \\
 \cline{2-6}
 & TopoRAG & \textbf{70.95$\pm$1.76} & \textbf{90.54$\pm$0.67} & \textbf{73.86$\pm$0.97} & \textbf{82.17$\pm$0.73} \\ 
 \hline
\multirow{4}{*}{\textbf{MLP}} & None & 42.03$\pm$0.38 & 73.85$\pm$0.92 & 49.58$\pm$0.15 & 77.43$\pm$0.51 \\
 & RD & 39.93$\pm$0.41 & 70.34$\pm$0.96 & 49.08$\pm$0.50 & 76.66$\pm$0.56 \\
 & Text & 47.57$\pm$0.73 & 72.82$\pm$0.83 & 51.32$\pm$0.51 & 77.63$\pm$0.63 \\
 \cline{2-6}
 & TopoRAG & \textbf{68.36$\pm$0.55} & \textbf{89.40$\pm$0.57} & \textbf{72.22$\pm$2.86} & \textbf{79.38$\pm$0.53} \\
 \Xhline{2\arrayrulewidth}
\end{tabular}
\label{tab-task}
\end{table}

\subsubsection{Task-oriented Evaluation}
In addition to conventional metrics, we also utilize task-oriented metrics to assess the quality of the generated texts. Specifically, we evaluate the performance of node classification and link prediction using the generated texts from different baselines. As shown in Table~\ref{tab-task}, TopoRAG consistently achieves the highest performance across all GNN backbones in both node classification and link prediction on the Cora and Pubmed datasets. This indicates that the texts generated by TopoRAG are more closely aligned with the main topics (node classification) and citations (link prediction) of their corresponding papers. Moreover, we observe that the performance improvement is even more pronounced when using MLP compared to GCN/SAGE. This is because GNN-based models inherently leverage neighborhood information to enhance context and hence compromise the augmenting effect caused by incorporating additional knowledge by Topo-RAG. Despite this inherent benefit, equipping them with TopoRAG still boosts the performance as TopoRAG additionally considers the potential of incorporating texts of non-neighboring nodes while the message-passing of GCN/SAGE only considers neighboring texts. In addition, we can see the additional benefit of TopoRAG on Cora is more than the one on Pubmed. We hypothesize that this is due to the higher correlation between textual similarity and topological similarity of Cora (0.2099) than the one of Pubmed (0.1039) in Figure~\ref{fig-extra-corr}. Notably, the superiority of our method is more pronounced in the results shown in Table~\ref{tab-task} compared to "ROUGE" and "Bert-F1" in Table~\ref{tab-main}. This suggests that employing complex, task-oriented evaluation metrics can reveal subtle distinctions in the quality of generated texts. Such an approach is particularly valuable in the current landscape of Large Language Models (LLMs), providing a nuanced means to quantify the effectiveness of text generation.

\subsection{Impact of Starting Words}
We explore the impact of increasing the number of starting words, ranging from 0 to 100, on the performance of text generation for the Cora/Book datasets. Our findings reveal that TopoRAG consistently outperforms other methods. This advantage becomes even more obvious when the number of initial words becomes less. This trend suggests that a reduced count of starting words offers less contextual information, thereby heightening the need for the additional information provided by the TopoRAG. Moreover, we observe an interesting trend with BLEU scores initially increasing and then decreasing on Cora, whereas BertScore-F1 shows a consistent upward trajectory. This pattern can be attributed to the inherent characteristics of these evaluation metrics. The BLEU score focuses on the overlap of exact words. As we provide more starting words, the length of the remaining part of the target sentence decreases. Consequently, in the latter stages, the likelihood that the generated words precisely match the few remaining words diminishes, leading to a drop in BLEU scores. On the other hand, BertScore-F1 evaluates semantic embedding matching, which does not rely on exact word overlap. Therefore, as the number of provided starting words increases, LLMs gain a better understanding of the general context of the target text. This enhanced contextual understanding facilitates the generation of text that is semantically more aligned, explaining the consistent improvement in BertScore-F1. The reason why we do not see this first-increasing and then-decreasing trend with BLEU score on Book is due to the generally longer lengths of their texts compared with Cora, as verified in Figure~\ref{fig-sent_emails}(b).

\begin{figure}[t!]
    \centering
    \includegraphics[width=0.48\textwidth]{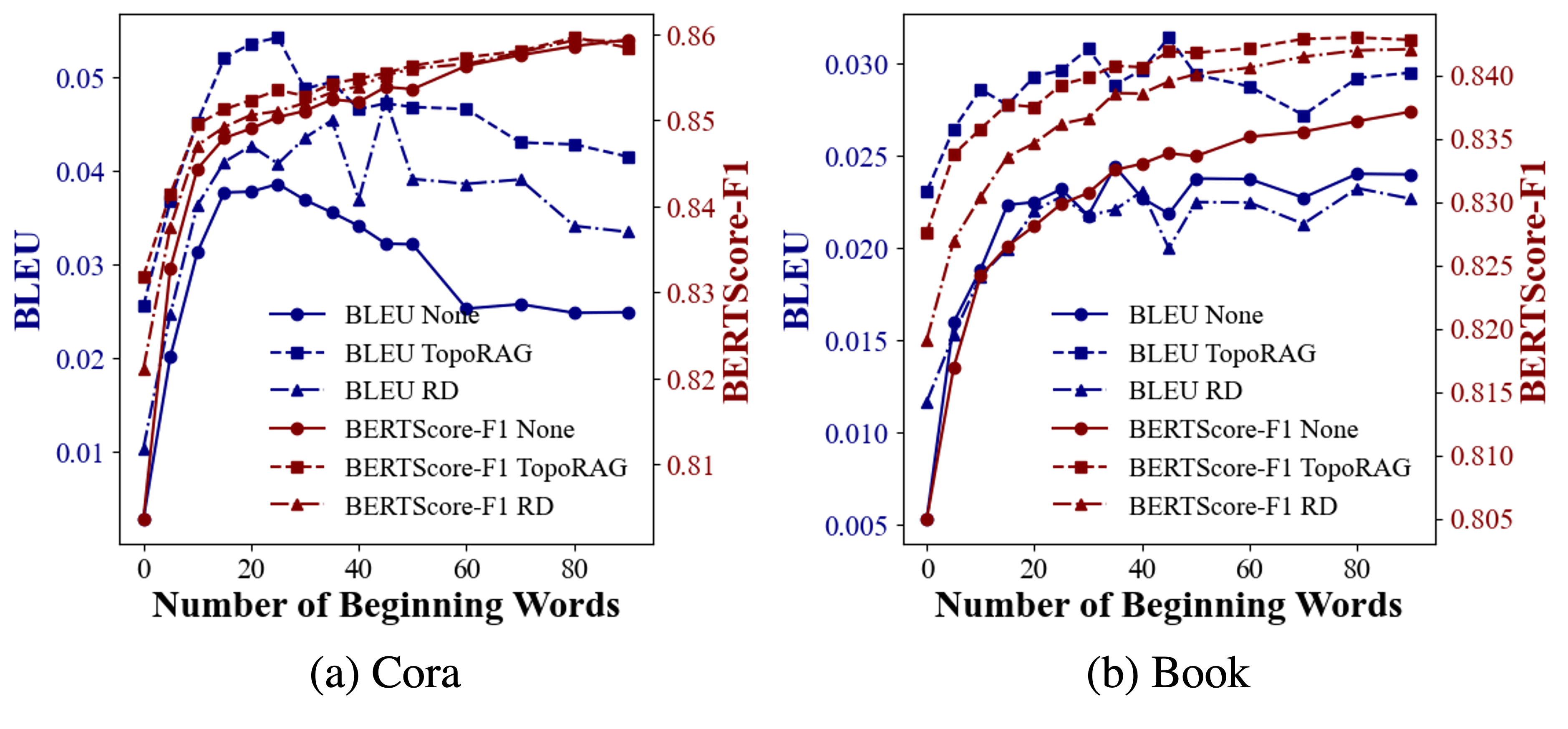}
    \vspace{-5ex}
    \caption{As the number of beginning words increases, the BLEU score first increases and then decreases on Cora while the BertScore-F1 continually increases. TopoRAG Consistently achieves the highest text generation performance than the other two baselines.}
    \label{fig-start-words}
    \vspace{-3ex}
\end{figure}

\subsection{Feature Imputation with TopoRAG}
Many machine learning models assume a fully observed feature matrix. However, in practice, each feature is only observed for a subset of nodes due to constraints like privacy concerns or limited resources for data annotation~\cite{yoon2018gain}. In all these scenarios, the missing feature issues could catastrophically compromise the capability of machine learning models~\cite{rossi2022unreasonable}, which motivates many previous works developing solutions to handling missing feature issue~\cite{you2020handling}. 

Since our proposed TopoRAG can naturally generate node features in graph-based datasets, in this section, we evaluate its effectiveness in handling missing features by comparing its performance against other conventional baselines handling missing node features for graph-based tasks. We take the Cora dataset and consider two tasks, node classification (NC) and link prediction (LP). For NC, we follow the traditional semi-supervised setting~\cite{wang2021tree} and for LP, we divide training/validation/testing links by 70\%/10\%/20\% following~\cite{zhao2022learning}. For baselines handling missing feature issues, we consider baselines that set missing features to 0 \textbf{(Zero)}; a random value from a standard Gaussian \textbf{(Random)}; and also the global mean of that feature over the graph \textbf{(Global Mean)}. We consider equipping the backbone GCN and MLP with each of these strategies. In Figure~\ref{fig-imputation-application}, we visualize the performance of different baselines at different missing feature rates from 0.1 to 0.9. We can observe that TopoRAG consistently outperforms other strategies in both node classification and link prediction across all rates of missing features. This underscores the benefits of incorporating additional context in handling missing feature issues on graphs.

\section{Related Work}\label{sec-relatedwork}

\subsection{Retrieval Augmented Generation}
Retrieval augmented generation (RAG) refers to augmenting the performance of generative-based tasks via retrieving relevant information from external knowledge bases~\cite{lewis2020retrieval, karpukhin2020dense, izacard2022few, li2022survey, feng2023trends, he2024g, gao2023retrieval}. Conventionally, RAG is widely used in enhancing the performance of question-answering tasks by retrieving supporting facts including the answer to the given question~\cite{karpukhin2020dense, xiong2020answering, ju2022grape, liu2023knowledge}. With the advent of LLMs, RAG gained more proliferation due to its capability to remedy the disadvantages of LLMs, such as mitigating the hallucination issue~\cite{zhang2023siren, yao2023llm, bang2023multitask}, enhancing interpretability~\cite{gao2023chat}, and enabling dynamic knowledge evolution of LLMs~\cite{wang2023knowledge, li2022parameter}. Significant research efforts have recently been devoted to improving RAG through developing better retrieval methods~\cite{wang2023knowledge, zhang2023graph, wang2023knowledgpt, trivedi2022interleaving} or incorporating knowledge bases of various modalities~\cite{kandpal2023large, kumar2023automatic}. Following the former research trend, we enhance RAG retrieval methods by equipping it with topology awareness. Different from KG-based RAG where topology information is incorporated by retrieving triples from subgraphs around entities mentioned in the question~\cite{wang2023knowledge}, we explicitly consider proximity and role-based topological relations in guiding the retrieval, the related works of which are reviewed next.

\subsection{Proximity/Role-based Topological Relations}
Real-world entities often exhibit interconnected relationships that can be classified into two primary categories: proximity-based and structural-based relationships~\cite{rossi2014role, rossi2020proximity, fortunato2010community, schaeffer2007graph, ahmed2018learning}. Proximity-based relationships between two nodes focus on their topological proximity, such as friends/relatives in social networks, co-cited academic papers in citation networks, and products co-purchased by the same customer~\cite{zhu2021node, garcia2017learning, hamilton2017representation, ahmed2018learning}. On the other hand, structural-based relations focus on the topological similarity between the local sub-structures of two nodes, e.g., two employees possessing the same title in a company share similar job responsibilities or airports acting as hubs following specific airline patterns~\cite{donnat2018learning, ribeiro2017struc2vec}. Previous works design various embedding-based methods capturing these two topological patterns. Methods such as Node2Vec and DeepWalk are mainly designed for capturing proximity-based topology patterns~\cite{grover2016node2vec, perozzi2014deepwalk} while Struc2Vec~\cite{ribeiro2017struc2vec} and GraphWave~\cite{donnat2018learning} are mainly to capture structure-based patterns. Different from them, we explore whether these topological patterns also correlate to the textual patterns, e.g., whether two closely interacted people share similar tweet contents in their Twitter accounts or two structurally similar employees share similar job descriptions. Furthermore, we leverage the discovered correlation to guide the retrieval and enhance text generation.

\section{Conclusion and Future Work}\label{sec-conclusion}
Given the limited knowledge provided in the input texts and the hallucination problem of LLMs, traditional approaches have leveraged RAG to incorporate extra knowledge. However, they predominantly focus on question-answering tasks with no significant investment in text-generation tasks. Furthermore, they overlook two critical types of knowledge embedded in the topological space, proximity-based and role-based knowledge. Therefore, this research aims to improve text generation performance by incorporating these two topological knowledge. Our empirical analysis reveals that LLMs benefit from additional texts that are similar to the target text. Moreover, by analyzing a wide range of text-attributed networks from diverse domains, we empirically verify the noticeable positive correlation between textual and proximity/role-based similarity. These findings have inspired us to develop Topology-aware Retrieval-Augmented Generation (Topo-RAG), a framework that enhances text generation by retrieving texts based on their topological similarities to the target text. We conduct comprehensive experiments to validate the effectiveness of Topo-RAG in text generation. Moreover, we take the initiative in utilizing node classification and link prediction to quantify the quality of the generated texts in a novel task-oriented manner. Additionally, we showcase an application of Topo-RAG in addressing missing feature issues in graph machine learning tasks.

Recognizing the importance of not only considering the quantity but also the structure of input knowledge in text generation~\cite{white2023prompt}, future work will focus on optimizing input formats by leveraging topological signals for question-answering and text-generation tasks. Moreover, we plan to assess the robustness of the TopoRAG framework by exploring the potential of attacking/defending over graphs to compromise/strengthen the capability of LLMs in completing downstream tasks.

\appendix
\section{Appendix}
\begin{figure}[t!]
    \centering
    \includegraphics[width=0.5\textwidth]{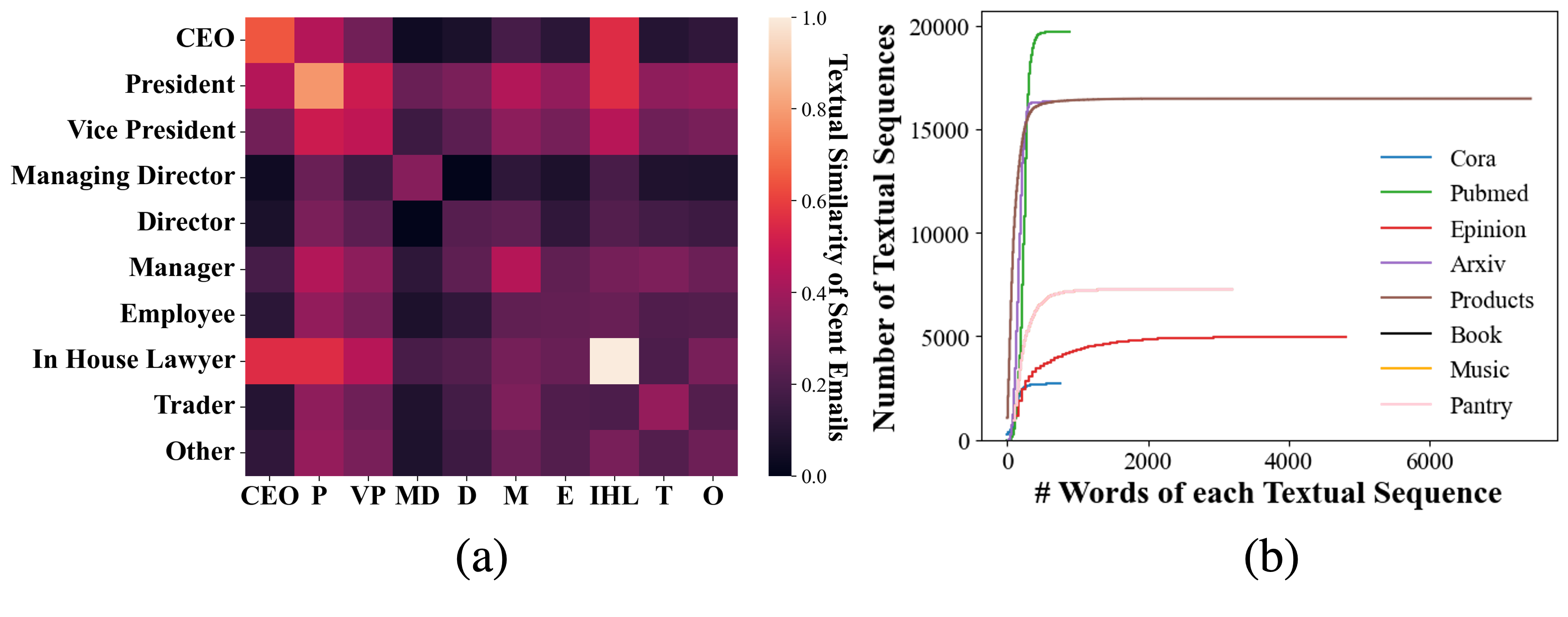}
    \caption{(a) The textual similarity of sent emails by pairs of employees grouped based on their job titles; (b) The distribution of passage length for each dataset.}
    \label{fig-sent_emails}
\end{figure}

\begin{figure}[t!]
    \centering    \includegraphics[width=0.5\textwidth]{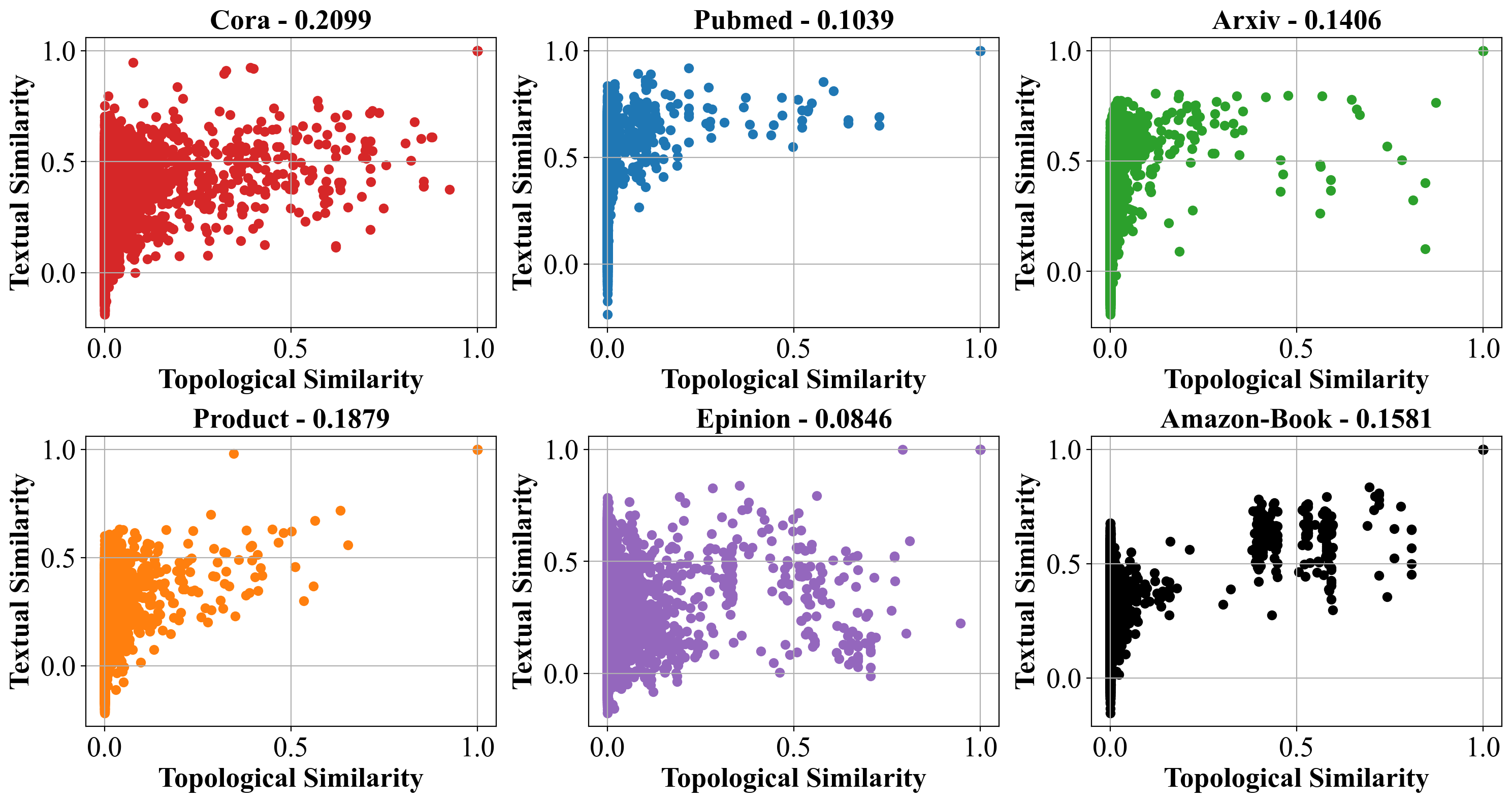}
    \caption{Correlation analysis between the textual similarity and proximity-based topological similarity over six datasets. The correlation shown beside the name of the dataset is consistently positive across different datasets from different domains.}
    \label{fig-extra-corr}
\end{figure}

In this section, we present supplementary findings that enhance the analysis from our primary study and further validate the generalizability of the observed phenomenon.

\subsection{Textual Similarity over Sent Emails on Eron-Email Dataset}\label{app-sent}
Following the same setting used for Figure~\ref{fig-motivation}(c), we visualize the textual similarity of emails sent by any pair of employees and further group them based on their job titles. Similar to the observation in Figure~\ref{fig-motivation}(c), we can see that people sharing the same job titles have similar textual patterns in their sent emails.

\subsection{Additional Correlation Analysis}\label{app-cor}
Here we conduct additional correlation analysis to demonstrate the positive correlation between proximity-based topological similarity and textual similarity. We can see a consistent positive correlation across six datasets from citation and E-commerce domains. This further justifies why TopoRAG achieves almost consistently higher performance than other baselines on all these datasets.


\bibliographystyle{ACM-Reference-Format}
\bibliography{reference}

\end{document}